%% file: main.tex
\documentclass{iopjournal}

\usepackage{amsmath,amssymb,bm}
\usepackage{booktabs}
\usepackage{array}
\usepackage{tabularx}
\usepackage{cite}
\usepackage{placeins}
\usepackage{flafter}
\usepackage{microtype}
\microtypesetup{expansion=false}

\makeatletter
\newif\ifcpcappendix
\cpcappendixfalse
\let\cpc@originalappendix\appendix
\renewcommand{\appendix}{%
  \cpc@originalappendix
  \global\cpcappendixtrue
  \@addtoreset{equation}{section}%
  \renewcommand{\theequation}{\Alph{section}\arabic{equation}}%
  \renewcommand{\thesubsection}{\thesection.\arabic{subsection}}%
  \renewcommand{\thesubsubsection}{\thesubsection.\arabic{subsubsection}}%
  \renewcommand\subsection{\@startsection{subsection}{2}{\z@}%
    {-2.2ex\@plus -0.6ex\@minus -0.2ex}%
    {0.7ex\@plus 0.2ex}%
    {\reset@font\normalsize\itshape}}%
}
\renewcommand{\thesection}{\Roman{section}}
\renewcommand{\thesubsection}{\thesection.\Alph{subsection}}
\renewcommand{\thesubsubsection}{\thesubsection.\arabic{subsubsection}}
\renewcommand{\@seccntformat}[1]{%
  \ifstrequal{#1}{section}{%
    \ifcpcappendix \Alph{section}.\enspace
    \else \Roman{section}.\enspace\fi
  }{\ifstrequal{#1}{subsection}{%
    \ifcpcappendix \arabic{subsection}.\enspace
    \else \Alph{subsection}.\enspace\fi
  }{\ifstrequal{#1}{subsubsection}{%
    \arabic{subsubsection}.\enspace
  }{%
    \csname the#1\endcsname\enspace
  }}}
}
\renewcommand\section{\@startsection{section}{1}{\z@}%
  {-3.0ex\@plus -0.8ex\@minus -0.2ex}%
  {1.2ex\@plus 0.2ex}%
  {\reset@font\normalsize\bfseries\centering}}
\renewcommand\subsection{\@startsection{subsection}{2}{\z@}%
  {-2.6ex\@plus -0.8ex\@minus -0.2ex}%
  {0.9ex\@plus 0.2ex}%
  {\reset@font\normalsize\bfseries\centering}}
\renewcommand\subsubsection{\@startsection{subsubsection}{3}{\z@}%
  {-2.2ex\@plus -0.6ex\@minus -0.2ex}%
  {0.7ex\@plus 0.2ex}%
  {\reset@font\normalsize\itshape}}
\makeatother

\let\cpcoldthebibliography\thebibliography
\let\cpcendoldthebibliography\endthebibliography
\renewenvironment{thebibliography}[1]{%
  \cpcoldthebibliography{#1}%
  \fontsize{9}{10.5}\selectfont
  \setlength{\itemsep}{1.5pt}%
  \setlength{\parsep}{0pt}%
}{\cpcendoldthebibliography}

\graphicspath{{figures/}{./}}
\newcommand{\orcidicon}{\raisebox{-0.15ex}{\includegraphics[height=1.4ex]{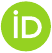}}}

\newcommand{\cpcfrontnotes}{%
  \begingroup
  \renewcommand{\thefootnote}{%
    \ifcase\value{footnote}\or *\or
    \ensuremath{\dagger}\or
    \ensuremath{\ddagger}\or
    \S\else
    \arabic{footnote}\fi}%
  \footnotetext[1]{This work was supported in part by the National Natural Science Foundation of China (NSFC) under Grant No.~12347147 (M.~Gao).}%
  \footnotetext[2]{Corresponding author. Email: \href{mailto:yuboliang@sjtu.edu.cn}{yuboliang@sjtu.edu.cn}}%
  \footnotetext[3]{Email: \href{mailto:SJTU-zy-zhouruixin@sjtu.edu.cn}{SJTU-zy-zhouruixin@sjtu.edu.cn}}%
  \footnotetext[4]{Email: \href{mailto:msgao@ecust.edu.cn}{msgao@ecust.edu.cn}}%
  \endgroup
}

\hypersetup{
  pdftitle={Entanglement spectrum ordering and flavor polarization in the two-flavor Schwinger model at vacuum angle theta=pi},
  pdfsubject={Cut-dependent gauge-sector ordering, Schmidt weights, and conditional flavor polarization in finite Schwinger chains},
  pdfkeywords={Schwinger model, matrix-product states, DMRG, symmetry-resolved entanglement, staggered fermions}
}

\begin{document}
\articletype{Paper}
\title{Entanglement spectrum ordering and flavor polarization in the two-flavor Schwinger model at vacuum angle $\theta=\pi$}
\author{Boliang Yu$^{1,3,\dagger}$, Ruixin Zhou$^{1,\ddagger}$ and Meisen Gao\,\href{https://orcid.org/0000-0002-4028-9895}{\orcidicon}$^{2,3,\S}$\par}
\affil{$^{1}$School of Physics and Astronomy, Shanghai Jiao Tong University, Shanghai 200240, China\\
$^{2}$School of Physics, East China University of Science and Technology, Shanghai 200237, China\\
$^{3}$Shanghai Key Laboratory of Particle Physics and Cosmology, Shanghai 200240, China\par}
\cpcfrontnotes

\begin{abstract}
Entanglement spectra in gauge theories can encode both symmetry breaking and the organization of gauge sectors. In the two-flavor Schwinger model at vacuum angle $\theta=\pi$, we find that the joint Schmidt distribution $p_{Q_A,F_A}$, labeled by the subsystem gauge charge $Q_A$ and flavor imbalance $F_A$, reveals a cut-dependent gauge-sector hierarchy and a mass-induced flavor asymmetry: changing the staggered cut reorganizes the gauge-charge distribution $p_{Q_A}$, while mass imbalance breaks the $F_A\leftrightarrow-F_A$ symmetry of the conditional flavor distribution $p_{F_A|Q_A}$. Defining the combined weight $W_F^{(q)}=p_{q,+1}+p_{q,-1}$ and conditional polarization $\mathcal P_F^{(q)}=(p_{q,+1}-p_{q,-1})/W_F^{(q)}$, we find that changing the cut reverses the weight hierarchy, $W_F^{(-1)}>W_F^{(+1)}$ at unit-cell boundaries but $W_F^{(+1)}>W_F^{(-1)}$ at intra-cell cuts, while mass imbalance drives $\mathcal P_F^{(q)}$ away from zero with a $q$-dependent cut response. Symmetry resolution therefore separates gauge-sector ordering, sector weight, and flavor polarization that are mixed in the globally ordered entanglement spectrum.
\end{abstract}

\input{sections/manuscript_body}

\bibliographystyle{unsrt}
\bibliography{references}
\end{document}

%% file: sections/manuscript_body.tex
\section{Introduction}
\label{sec:intro}

The Schwinger model is a compact setting for studying confinement, background electric flux, chiral structure, and vacuum-angle dependence with both field-theoretical and nonperturbative methods: the massless theory is exactly solvable and generates a massive gauge boson through the anomaly \cite{schwinger1962}, while the massive theory acquires a nontrivial dependence on the vacuum angle $\theta$ \cite{coleman1976}. Its multi-flavor extensions add flavor dynamics and explicit mass imbalance while retaining a one-dimensional formulation well suited to tensor-network calculations \cite{hetrick1995,dempsey2024,albandea2025,kanno2025,cuomo2026}. The region around $\theta=\pi$ is especially subtle because competing vacuum structures and long correlation lengths amplify finite-size and boundary effects.

Because the Kogut--Susskind Hamiltonian lattice formulation \cite{kogut1975} makes the model directly accessible to real-space variational methods, it has become a standard proving ground for tensor networks in lattice gauge theory. Density-matrix renormalization group (DMRG) calculations resolved the mass spectrum of the one-flavor massive model two decades ago \cite{byrnes2002}, and matrix-product-state (MPS) formulations that either eliminate the gauge links by Gauss's law or retain them explicitly were developed shortly afterwards \cite{banuls2013,buyens2014,rico2014}. The approach has since produced the chiral condensate at finite lattice spacing \cite{banuls2016condensate}, the confining string and its breaking \cite{buyens2016}, density-induced transitions \cite{banuls2017density}, the topological vacuum structure at finite $\theta$ \cite{funcke2020theta}, the CP-violating Dashen phase \cite{funcke2023dashen}, and combined analytic and DMRG results on a finite interval \cite{okuda2023}; broader accounts of the method and its reach are given in Refs.~\cite{dalmonte2016,banuls2019lgt,banuls2020review}. More recently, gauge-invariant uniform MPS have been used to characterize the critical behavior of the one-flavor model \cite{fujii2025}.

Quantities computed from the entanglement structure of these states have become observables in their own right. The entanglement entropy of a one-dimensional critical system is controlled by the central charge \cite{calabrese2004}, while the full Schmidt spectrum, or entanglement spectrum, retains structural information that the entropy alone discards \cite{li2008}. When the state carries a global symmetry the reduced density matrix decomposes into charge sectors; the sector weights can equivalently be viewed as subsystem-charge probabilities generated by charged moments, linking symmetry-resolved entanglement to full counting statistics \cite{goldstein2018,murciano2020,bonsignori2019}. This structure has also been shown to be sensitive to open boundaries \cite{bonsignori2021}. In a gauge theory the subsystem algebra and the Hilbert-space factorization require an explicit prescription \cite{casini2014}, which the superselection-resolved tensor-network construction of Ref.~\cite{feldman2024superselection} supplies for lattice gauge theories. In the Schwinger model itself, entanglement entropy and a global entanglement-spectrum spacing have been used to diagnose the one-flavor critical region \cite{ikeda2023entanglement}, and finite-$\theta$ work relates entropy enhancement to competition between electric-flux branches \cite{grieninger2026theta}.

The two-flavor theory supplies the flavor structure that the present work exploits. Hetrick, Hosotani and Iso established the bosonized description of the massive multi-flavor model \cite{hetrick1995}, and Smilga obtained the critical amplitudes that organize its small-mass expansion \cite{smilga1997}. Dempsey \textit{et al.} mapped the zero-temperature phase diagram and identified the role of staggered-lattice charge conjugation at $\theta=\pi$ \cite{dempsey2024}, a line continued in Ref.~\cite{cuomo2026}, where the strong-coupling low-energy description at $\theta=\pi$ is developed in detail. On the numerical side, DMRG has resolved the $\theta$-dependent meson spectrum \cite{itou2024dmrg,matsumoto2025}, MPS calculations have determined the additive mass shift of Wilson and staggered fermions \cite{angelides2023} and compared fermion discretizations directly \cite{schwaegerl2025discretization}, Grassmann tensor-renormalization-group calculations have resolved thermodynamic-limit lattice phase structure with a $\theta$ term \cite{kanno2025}, and lattice Monte Carlo has addressed chiral and isospin breaking \cite{albandea2025}.

The diagnostic of Ref.~\cite{ikeda2023entanglement} is constructed from the unresolved spectrum, whose globally leading levels need not belong to the same symmetry block. The central physical question is therefore whether a change in the global ordering reflects competition between gauge-flux sectors or a flavor response within a fixed sector. We address this question through the joint Schmidt-probability distribution $p_{Q_A,F_A}$: its marginal $p_{Q_A}$ records the gauge-charge organization, while $p_{F_A\mid Q_A}$ resolves the flavor distribution at fixed subsystem charge. For the complementary $(Q_A,F_A)=(\pm1,\pm1)$ pairs, we distinguish the combined pair weight $W_F^{(q)}$ from the pair-normalized conditional polarization $\mathcal P_F^{(q)}$, while independently tracking the gauge-sector identity of the globally leading levels across the two staggered cut classes. This separates three quantities that an unresolved entanglement spectrum mixes together: the gauge-charge marginal, the $F_A=\pm1$ pair weight and flavor asymmetry within a fixed charge sector, and the sector's position in the global Schmidt ordering.

We find that changing the staggered cut class switches the dominant non-neutral Schmidt pair from $Q_A=-1$ at unit-cell boundaries to the complementary $Q_A=+1$ pair at intra-cell cuts. Mass imbalance drives both $\mathcal P_F^{(-1)}$ and $\mathcal P_F^{(+1)}$ away from zero with a charge- and cut-dependent response: at unequal mass, $\mathcal P_F^{(-1)}$ is positive for both cut classes, whereas $\mathcal P_F^{(+1)}$ is negative at unit-cell boundaries and positive intra-cell. Within the tracked $Q_A=-1$ pair, mass imbalance also increases its combined probability, even though the intra-cell pair weight is strongly suppressed relative to the unit-cell value. The complete joint distribution further distinguishes non-neutral broadening at unit-cell boundaries from redistribution among non-neutral sectors at intra-cell cuts. The associated entanglement-level difference $D_F$ is retained as a secondary logarithmic representation of $\mathcal P_F^{(-1)}$ rather than a measure of normalized-block restructuring. We establish these results using exact-diagonalization-validated matrix-product-state calculations and symmetry-resolved Schmidt decompositions.

This paper is organized as follows. Section~\ref{sec:framework} introduces the lattice formulation, subsystem quantum numbers, Schmidt-sector probabilities, and numerical methods. Section~\ref{sec:results} presents the numerical validation, cut-dependent gauge-sector ordering, joint probability distribution, flavor-pair response, and finite-chain controls. Section~\ref{sec:discussion} discusses the physical content and scope, and Sec.~\ref{sec:summary} summarizes the results. Formal atomic-reference selection rules, logarithmic fit diagnostics, convergence checks, and cut-selection controls are provided in the Appendix, while the availability of the underlying datasets is stated in the Data availability statement.

\section{Model and symmetry-resolved Schmidt probabilities}
\label{sec:framework}

Detailed accounts of the Hamiltonian lattice formulation and of its matrix-product-state implementation can be found in Refs.~\cite{kogut1975,banuls2013,schollwoeck2011}. Here we reproduce only the conventions on which the later analysis depends, and then state the numerical settings in full.

\subsection{Hamiltonian and open-boundary conventions}
\label{sec:model}

We use the Kogut--Susskind Hamiltonian lattice formulation \cite{kogut1975} for a two-flavor Schwinger model with \(N_s\) matter sites. Matter sites are numbered \(n=0,\ldots,N_s-1\), and each physical site contains two fermionic modes labeled by \(\alpha=1,2\). The gauge field is eliminated using Gauss's law under open boundary conditions. We fix a bare link flux \(L_{-1}\) to the left of the chain and reconstruct the internal electric fields using Eq.~\eqref{eq:gauss-law}:

\begin{equation}
L_n-L_{n-1}=Q_n,
\qquad
L_n=L_{-1}+\sum_{k=0}^{n}Q_k.
\label{eq:gauss-law}
\end{equation}

Equations~\eqref{eq:hamiltonian-sum}--\eqref{eq:electric} define the Hamiltonian used in all three numerical representations. The Hamiltonian includes the \(N_s-1\) internal dynamical links \(n=0,\ldots,N_s-2\). The right external flux is determined by the total charge,

\begin{equation}
L_{\mathrm{right}}
=
L_{-1}+Q_{\mathrm{tot}},
\end{equation}

but its energy is not included in the Hamiltonian used for the quantitative results. This distinction is relevant away from a fixed total-charge sector. The internal-link convention is used throughout this work.

The Hamiltonian is decomposed as

\begin{equation}
H=H_{\mathrm{hop}}+H_{\mathrm{mass}}+H_E.
\label{eq:hamiltonian-sum}
\end{equation}

\begin{equation}
H_{\mathrm{hop}}
=
-\frac{i}{2a}
\sum_{n=0}^{N_s-2}\sum_{\alpha=1}^{2}
\left(
c^\dagger_{n,\alpha}c_{n+1,\alpha}
-
c^\dagger_{n+1,\alpha}c_{n,\alpha}
\right),
\label{eq:hopping}
\end{equation}

\begin{equation}
H_{\mathrm{mass}}
=
\sum_{n=0}^{N_s-1}\sum_{\alpha=1}^{2}
m_{\mathrm{lat},\alpha}(-1)^n n_{n,\alpha},
\qquad
m_{\mathrm{lat},\alpha}
=
m_\alpha-\frac{g^2a}{4},
\label{eq:mass}
\end{equation}

and

\begin{equation}
H_E
=
\frac{g^2a}{2}
\sum_{n=0}^{N_s-2}
\left[
L_{-1}
+\sum_{k=0}^{n}Q_k
+\frac{\theta}{2\pi}
\right]^2.
\label{eq:electric}
\end{equation}

The local gauge charge appearing in Eq.~\eqref{eq:gauss-law}, including the staggered background of both flavors, is

\begin{equation}
Q_n
=
\sum_{\alpha=1}^{2}
\left[
n_{n,\alpha}
-
\frac{1-(-1)^n}{2}
\right]
=
n_{n,1}+n_{n,2}-\left[1-(-1)^n\right].
\label{eq:site-charge}
\end{equation}

The definition in Eq.~\eqref{eq:site-charge} includes the staggered background; the background angle shifts the electric energy but does not enter Gauss's law or \(Q_n\). The total conserved quantum numbers used to specify the canonical sector are

\begin{equation}
Q_{\mathrm{tot}}=\sum_{n=0}^{N_s-1}Q_n,
\qquad
F_{\mathrm{tot}}=\sum_{n=0}^{N_s-1}(n_{n,1}-n_{n,2}).
\end{equation}

We use \(g\) as the unit of mass and energy and set \(g=1\). Throughout the quantitative calculations,

\begin{equation}
\theta=\pi,\qquad L_{-1}=0,
\end{equation}

and the MPS is restricted to

\begin{equation}
Q_{\mathrm{tot}}=0,
\qquad
F_{\mathrm{tot}}=0.
\end{equation}

We work in the zero-charge, flavor-neutral canonical sector. Every response MPS is the lowest-energy state obtained in this sector and satisfies the criteria of Sec.~\ref{sec:dmrg}. The exact symmetry maps in Appendix~\ref{app:validation} provide additional checks of the sector construction. Staggered-lattice charge conjugation also translates matter and links by one site, schematically $c_{n,\alpha}\mapsto c^\dagger_{n+1,\alpha}$ and $L_n\mapsto-L_{n+1}-1$ at $\theta=\pi$ \cite{dempsey2024}. The choice $L_{-1}=0$, together with the open termination and internal-link electric-energy convention, therefore defines a finite-chain realization that is not invariant under the staggered charge-conjugation transformation. In the $C$-broken region of the infinite-volume phase diagram, this boundary condition removes the finite-system equivalence of the two $C$-related bulk vacua and selects one finite-chain continuation; we do not infer from this selection the vacuum degeneracy of the thermodynamic system.

The dimensionless lattice parameter is

\begin{equation}
x=\frac{1}{g^2a^2}.
\label{eq:x-definition}
\end{equation}

The gauge-covariant staggered Hamiltonians of Refs.~\cite{dempsey2024,cuomo2026} use the same imaginary antisymmetric hopping, Gauss-law charge, and two-flavor mass shift; the latter has also been determined with MPS methods \cite{angelides2023}. On the open chain we choose link phases $U_n=1$ and the internal-link endpoint convention of the gauge-eliminated OBC formulation \cite{banuls2013}. The operator-independent constant in the shifted electric squares is retained, while the staggered background is already included in $Q_n$. Appendix~\ref{app:electric-mpo} gives the exact electric-term rearrangement and MPO coefficients.

\subsection{Charges, cut classes, and flavor exchange}
\label{sec:local-basis}

\subsubsection{Local basis and fermionic ordering}

The local Hilbert space has dimension four. We use

\begin{equation}
|n_1,n_2\rangle
=
|0,0\rangle,\ |1,0\rangle,\ |0,1\rangle,\ |1,1\rangle.
\end{equation}

The intra-site fermionic sign is fixed so that the two local flavors satisfy the canonical anticommutation relations. The full-chain mode ordering is site-major,

\begin{equation}
(0,1),(0,2),(1,1),(1,2),\ldots,(N_s-1,1),(N_s-1,2).
\end{equation}

Cross-site operators use the corresponding Jordan--Wigner strings, with the same site-major ordering in ED and the tensor-network implementation.

\subsubsection{Subsystem charges}
\label{sec:subsystem-charges}

For a cut after the first \(\ell\) physical sites, the left subsystem is

\begin{equation}
A=\{0,\ldots,\ell-1\}.
\end{equation}

We define

\begin{equation}
N_{1,A}=\sum_{n<\ell}n_{n,1},
\qquad
N_{2,A}=\sum_{n<\ell}n_{n,2},
\end{equation}

with

\begin{equation}
N_A=N_{1,A}+N_{2,A},
\end{equation}

and the integer flavor imbalance

\begin{equation}
F_A=N_{1,A}-N_{2,A}.
\end{equation}

The physical subsystem gauge charge is

\begin{equation}
Q_A=\sum_{n<\ell}Q_n.
\end{equation}

The staggered background contribution is included in \(Q_A\); it cannot be replaced by the bare occupation number. Gauss's law gives \(L_{\ell-1}=L_{-1}+Q_A\). Thus \(Q_A\) labels the electric-flux superselection sector across the cut, while \(F_A\) further resolves the conserved flavor-\(U(1)\) block \cite{feldman2024superselection}. In terms of the subsystem occupation number,

\begin{equation}
Q_A=N_A-2\left\lfloor\frac{\ell}{2}\right\rfloor.
\label{eq:subsystem-charge-map}
\end{equation}

Appendix~\ref{app:charge-map} gives the tensor-network label mapping and its ED validation.

\subsubsection{Staggered unit cells and cut classes}

The bipartition itself is not neutral with respect to the staggered structure, and the classification that follows is used throughout the analysis. With the chosen staggered convention, two adjacent matter sites form a natural unit cell,

\begin{equation}
(2j,2j+1).
\end{equation}

A cut \(\ell\) is defined between sites \(\ell-1\) and \(\ell\). Odd \(\ell\) and even \(\ell\) belong to different bond classes: one lies inside the staggered unit cell, while the other lies between complete unit cells. On an even-\(N_s\) chain, \(N_s\bmod4\) fixes the central-cut class. The central cut is a unit-cell boundary for \(N_s\bmod4=0\) and an intra-cell cut for \(N_s\bmod4=2\). The deterministic rule that selects the analysis cut among the available unit-cell boundaries is given in Appendix~\ref{app:cut-classification}.

Figure~\ref{fig:conventions} summarizes the boundary, local-basis, fermionic-ordering, and cut conventions used below.

\begin{figure}[!t]
\centering
\includegraphics[width=\textwidth]{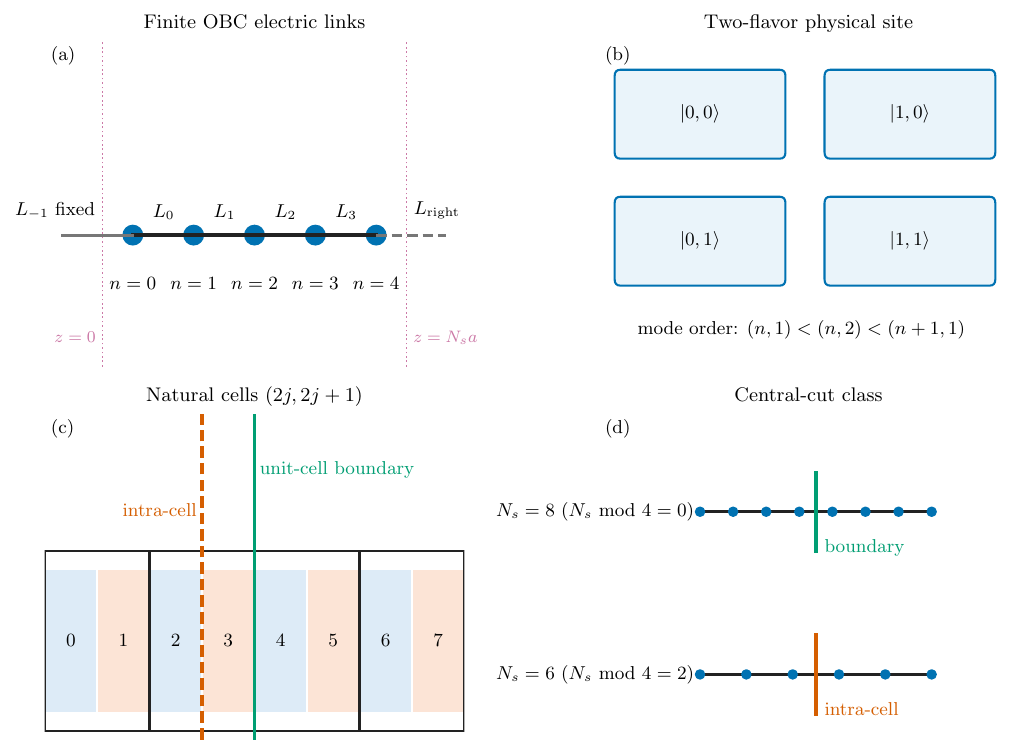}
\caption{Open-boundary link convention, the four-state two-flavor local basis, site-major fermion ordering, the natural staggered unit cell $(2j,2j+1)$, and the intra-cell and unit-cell-boundary cut classes. The symbols $n$, $j$, and $\ell$ are integer lattice indices; $a$ is the lattice spacing, and $z$ is a length coordinate with the chain endpoints shown as $0$ and $N_sa$.}
\label{fig:conventions}
\end{figure}

\subsection{Mass-ratio parameterization and flavor exchange}
\label{sec:mass-ratio}

We fix the geometric mean mass and vary the ratio

\begin{equation}
m_{\mathrm{geom}}=\sqrt{m_1m_2},
\qquad
r=\frac{m_1}{m_2},
\label{eq:mass-ratio-parameterization}
\end{equation}

so that

\begin{equation}
m_1=m_{\mathrm{geom}}\sqrt r,
\qquad
m_2=\frac{m_{\mathrm{geom}}}{\sqrt r}.
\end{equation}

The main parameter grid uses \(m_{\mathrm{geom}}/g=0.5\) and \(1\le r\le2\), from equal to nondegenerate masses. Along this path, \(m_1/g\) ranges from 0.5 to 0.707 and \(m_2/g\) from 0.5 to 0.354, resolving a mass-imbalance response without an extreme hierarchy. The geometric-mean constraint also makes exact reciprocal checks available because exchanging the two flavor labels maps \(r\) to \(\frac{1}{r}\). According to the zero-temperature phase diagram of Ref.~\cite{dempsey2024}, the sampled path $m_{\mathrm{geom}}/g=0.5$, $1\le r\le2$, lies within the region where charge conjugation $C$ is spontaneously broken in the infinite-volume theory. The equal-mass point lies on the $SU(2)$-invariant line inside this region, while increasing $r$ moves the path toward the Ising boundary without reaching it over the interval studied here. The selected interval therefore probes flavor imbalance within a single bulk phase rather than crossing a phase boundary, and it is not a pure flavor-breaking susceptibility path. Hence a flavor-invariant observable must obey

\begin{equation}
O(r)=O\!\left(\frac{1}{r}\right).
\end{equation}

It is useful to introduce

\begin{equation}
\eta=\ln r.
\end{equation}

The parameterization in Eq.~\eqref{eq:mass-ratio-parameterization} makes flavor exchange \(\eta\rightarrow-\eta\). Near equal mass, a smooth flavor-invariant response is therefore even in \(\eta\),

\begin{equation}
\delta O(\eta)
=
c_2\eta^2+c_4\eta^4+\cdots.
\end{equation}

Because flavor exchange is an on-site internal unitary, it factorizes across any physical-site bipartition as $\mathcal X=\mathcal X_A\otimes\mathcal X_{\bar A}$. It therefore preserves the Schmidt values while mapping $F_A\to-F_A$. For correspondingly exchanged ground states, with block rank (k) fixed after blockwise sorting,

\begin{equation}
\epsilon_k^{(Q_A,F_A)}(r)=\epsilon_k^{(Q_A,-F_A)}\!\left(\frac{1}{r}\right).
\label{eq:schmidt-flavor-exchange}
\end{equation}

We use these identities as analytical constraints and numerical consistency checks. The even expansion is descriptive only; no continuum response coefficient is extracted.

\subsection{Schmidt-sector probabilities}
\label{sec:sector-probabilities}

At a physical-site cut, the MPS Schmidt singular values $s_\alpha$ define the probabilities and entanglement energies
\begin{equation}
\lambda_\alpha=s_\alpha^2,
\qquad
\epsilon_\alpha=-\ln\lambda_\alpha.
\label{eq:schmidt-definitions}
\end{equation}
The symmetry blocks admit the normalized direct-sum decomposition
\begin{equation}
\rho_A=\bigoplus_{Q_A,F_A}p_{Q_A,F_A}\,\widetilde\rho_{Q_A,F_A},
\qquad
\operatorname{Tr}\widetilde\rho_{Q_A,F_A}=1,
\label{eq:rho-block-decomposition}
\end{equation}
where $p_{q,f}\equiv p_{Q_A=q,F_A=f}=\operatorname{Tr}\rho_A^{(q,f)}$ is the total Schmidt probability of the block labeled by the values $(q,f)$.
Equivalently,
\begin{equation}
p_{q,f}=\langle\Psi|\Pi^A_{q,f}|\Psi\rangle,
\label{eq:block-projector-probability}
\end{equation}
so the block weights form the normalized joint subsystem-charge distribution of $Q_A$ and $F_A$. We denote its gauge-charge marginal by $p_{Q_A}(q)\equiv\sum_f p_{q,f}$ and, whenever $p_{Q_A}(q)>0$, its conditional flavor distribution by $p_{F_A\mid Q_A}(f\mid q)\equiv p_{q,f}/p_{Q_A}(q)$. For these two commuting labels, define
\begin{equation}
Z_{1,A}(\alpha,\beta)
=\operatorname{Tr}\!\left[\rho_A e^{i\alpha Q_A+i\beta F_A}\right].
\label{eq:joint-charge-characteristic-function}
\end{equation}
Here $Z_{1,A}$ is the joint characteristic function, or $n=1$ charged moment, and its Fourier coefficients are $p_{q,f}$; higher charged moments generate symmetry-resolved R\'enyi moments. This is the two-charge extension of the charged-moment/full-counting-statistics construction used in symmetry-resolved entanglement \cite{goldstein2018,bonsignori2019,murciano2020}; the Schmidt decomposition additionally supplies the normalized spectrum within each block. For the complementary $q=\pm1$ flavor pairs, define
\begin{equation}
W_F^{(q)}=p_{q,+1}+p_{q,-1},
\qquad
\mathcal P_F^{(q)}
=\frac{p_{q,+1}-p_{q,-1}}{p_{q,+1}+p_{q,-1}},
\qquad q=\pm1.
\label{eq:pair-sum-polarization}
\end{equation}
Here $W_F^{(q)}$ is the absolute joint probability carried by the $F_A=\pm1$ pair within charge sector $q$. Because the common marginal cancels, $\mathcal P_F^{(q)}$ is equivalently the normalized contrast of $p_{F_A\mid Q_A}(\pm1\mid q)$; it is not, when other flavor values are present, the first moment of the complete conditional distribution. For the tracked $q=-1$ pair used in the derived entanglement-energy diagnostic, we further write
\begin{equation}
p_+\equiv p_{-1,+1},
\qquad
p_-\equiv p_{-1,-1}.
\label{eq:target-pair-weights}
\end{equation}

The lowest full-spectrum eigenvalue in each target block can be written as $\lambda_{1,\pm}=p_\pm\widetilde\lambda_{1,\pm}$. The block-minimum difference used below only as a derived logarithmic diagnostic is therefore
\begin{equation}
\begin{aligned}
D_F
&=\epsilon_1^{(-1,+1)}-\epsilon_1^{(-1,-1)}\\
&=\underbrace{\ln\!\left(\frac{p_-}{p_+}\right)}_{D_F^{\mathrm{weight}}}
+\underbrace{\ln\!\left(\frac{\widetilde\lambda_{1,-}}{\widetilde\lambda_{1,+}}\right)}_{D_F^{\mathrm{conf}}},\\
D_F^{\mathrm{weight}}
&=-2\operatorname{artanh}\!\left(\mathcal P_F^{(-1)}\right).
\end{aligned}
\label{eq:df-derived-decomposition}
\end{equation}
Thus $D_F^{\mathrm{weight}}$ is simply a monotone logarithmic representation of the conditional polarization. The smaller $D_F^{\mathrm{conf}}$ compares only the largest eigenvalues of the two normalized blocks and does not characterize their complete normalized spectra.

Flavor exchange gives exact parity constraints. With $\eta=\ln r$,
\begin{equation}
W_F^{(q)}(\eta)=W_F^{(q)}(-\eta),
\qquad
\mathcal P_F^{(q)}(\eta)=-\mathcal P_F^{(q)}(-\eta),
\qquad q=\pm1.
\label{eq:probability-flavor-parity}
\end{equation}
In particular, $\mathcal P_F^{(q)}(0)=0$ at equal mass. These constraints fix neither the nonzero magnitude nor the cut dependence. The probabilities, polarizations, entanglement energies, and their differences are dimensionless; entanglement energy is not a physical excitation energy made dimensionless by setting $g=1$. Throughout, $\ln$ denotes the natural logarithm.

The normalization of the complete Schmidt spectrum is checked before any low-lying-level filtering. All joint block probabilities $p_{q,f}$ used here are summed over every Schmidt value present in the available MPS state. The threshold $\lambda_\alpha>10^{-12}$ applies only to low-lying level tables and dominant-block identification; it is not applied to the block-weight reconstruction. Symmetry-resolved entanglement decompositions have been developed in continuum and lattice settings \cite{murciano2020,bonsignori2019}. In gauge theories the subsystem algebra and Hilbert-space factorization require an explicit prescription \cite{casini2014}; here the gauge links have first been eliminated by Gauss law, and the Schmidt cut is taken only between complete matter sites.

\subsection{Numerical methods and parameter coverage}
\label{sec:numerical-workflow}

\subsubsection{Sparse exact diagonalization and exact MPO}
\label{sec:ed-mpo}

Sparse ED provides the small-system reference. The mass and electric diagonal entries call the same local-charge and Gauss-law definitions used by the tensor-network construction; no independent boundary convention is introduced. Lowest eigenpairs are obtained with a sparse Hermitian eigensolver. Technical details, including explicit basis maps and the very-small-matrix fallback, are summarized in Appendix~\ref{app:validation}.

After summing over internal links, the site-dependent one-body and long-range two-body terms admit an exact finite-state MPO without an approximate long-range fit. Its construction and bond dimensions are given in Appendix~\ref{app:electric-mpo}.

\subsubsection{Finite-system DMRG and convergence}
\label{sec:dmrg}

We use TeNPy 1.1.0 \cite{hauschild2018,hauschild2024} with one four-state MPS site per physical matter site. The backend conserves the Abelian pair \((N,F)\), which is mapped to the physical subsystem labels \((Q_A,F_A)\) as stated in Sec.~\ref{sec:subsystem-charges}. The physical Hilbert basis and Hamiltonian are unchanged when these symmetries are enabled.

Finite-system DMRG \cite{white1992,schollwoeck2011} uses deterministic initialization, increasing bond dimension, and warm starts between successive \(\chi\). We define the energy variance per site as

\begin{equation}
v_E=\frac{\langle H^2\rangle-\langle H\rangle^2}{N_s}.
\end{equation}

For the main quantitative analysis, \(v_E\) and the maximum discarded weight must be below \(10^{-6}\), and the changes between the two largest available bond dimensions must satisfy

\begin{equation}
|\Delta E|<10^{-6},\qquad
\max_{k\le20}|\Delta\epsilon_k|<10^{-3},\qquad
\Delta O_{\mathrm{bulk}}<10^{-3}.
\end{equation}

The auxiliary bare-link-flux and staggered-occupation observables enter the main analysis only through the convergence monitor $\Delta O_{\mathrm{bulk}}$. Its deterministic bulk-window construction and the complete definitions of $L_{\mathrm{bulk}}$, $C_\pm^{\mathrm{bulk}}$, and $\Delta O_{\mathrm{bulk}}$ are given in Appendix~\ref{app:bulk-window}.

Here $\Delta E$ is the change in the canonical-sector ground-state energy between the two largest available bond dimensions. Entanglement levels are matched by $(Q_A,F_A)$ and block rank before the largest drift among the first 20 matched levels is taken. State normalization, canonical sector, Hamiltonian parameters, and cut definition are verified independently of the numerical thresholds. Reaching the prescribed maximum bond dimension is not by itself taken as evidence of convergence; convergence is assessed from the energy, spectrum, local-observable, and discarded-weight criteria stated above and reported in Sec.~\ref{sec:benchmarks}.

\subsubsection{Parameter coverage and numerical settings}
\label{sec:coverage}

We use \(g\) as the unit of mass and energy. The fixed physical parameters are

\begin{equation}
\theta=\pi,\qquad
m_{\mathrm{geom}}/g=0.5.
\end{equation}

The combined data set contains $x=4,6.25,9,12.25$ and a six-ratio scan $r=1,1.125,1.25,1.5,1.75,2$ at $gL_{\mathrm{phys}}\simeq8$. We write the physical spatial length as

\begin{equation}
L_{\mathrm{phys}}=N_s a,
\qquad
gL_{\mathrm{phys}}=\frac{N_s}{\sqrt{x}}.
\end{equation}

We place matter site \(n\) at \(z_n=\left(n+\frac{1}{2}\right)a\) between the geometric boundaries \(z=0\) and \(z=N_sa\); Appendix~\ref{app:coordinate-convention} records the full coordinate convention.

The three families satisfy \(gL_{\mathrm{phys}}\approx8,10,12\). They do not form a complete \(4\times3\times6\) Cartesian scan. The largest-length calculations contain only \(x=9,12.25\) and \(r=1,1.5,2\):

\begin{equation}
(x,N_s,gL_{\mathrm{phys}})=(9,36,12)
\end{equation}

and

\begin{equation}
(x,N_s,gL_{\mathrm{phys}})=(12.25,40,11.43),
\end{equation}

where the second length differs from its target by 4.76\%. Both have \(N_s\bmod4=0\), so the central cut is a unit-cell boundary.

The default comparison uses \(\chi=128,256\), and two matched \(r=1,1.5,2\) triplets were extended to \(\chi=384\). A response is retained only when it and its matched reference satisfy the convergence criteria. Final bond dimensions are listed in Table~\ref{tab:coverage}, with the exceptional drift details in Appendix~\ref{app:bulk-window}.

\begin{table}[!htbp]
\caption{Parameter coverage for the relative-response analysis. Every listed analysis cut is the central unit-cell boundary and every system has $N_s\bmod4=0$. The $gL_{\mathrm{phys}}\simeq10,12$ and near-equal calculations are targeted checks rather than a complete Cartesian scan. Matched $r=1$ references are shared where the numerical parameters coincide. Here $x$, $r$, $gL_{\mathrm{phys}}$, and $\chi$ are dimensionless; $N_s$ is the number of matter sites and $\ell_\star$ is an integer cut index.}
\label{tab:coverage}
\centering
\scriptsize
\begin{tabular*}{\textwidth}{@{\extracolsep{\fill}}l r r r l r r@{}}
\toprule
Subset & $x$ & $N_s$ & $gL_{\mathrm{phys}}$ & Mass ratios & $\chi$ & $\ell_\star$ \\
\midrule
$gL_{\mathrm{phys}}\simeq8$ scan & 4 & 16 & 8.00 & $1,1.125,1.25,1.5,1.75,2$ & 256 & 8 \\
 & 6.25 & 20 & 8.00 & same & 256 & 10 \\
 & 9 & 24 & 8.00 & same & 256 & 12 \\
 & 12.25 & 28 & 8.00 & same & 256 & 14 \\
\addlinespace
$gL_{\mathrm{phys}}\simeq10$ checks & 4 & 20 & 10.00 & $1,1.5,2$ & 256 & 10 \\
 & 6.25 & 24 & 9.60 & same & 256 & 12 \\
 & 9 & 28 & 9.33 & same & 256 & 14 \\
 & 12.25 & 36 & 10.29 & same & 384 & 18 \\
\addlinespace
$gL_{\mathrm{phys}}\simeq12$ checks & 9 & 36 & 12.00 & $1,1.5,2$ & 256 & 18 \\
 & 12.25 & 40 & 11.43 & same & 384 & 20 \\
\addlinespace
\bottomrule
\end{tabular*}
\end{table}

\FloatBarrier

\section{Results}
\label{sec:results}

Section~\ref{sec:benchmarks} summarizes the numerical benchmarks, Sec.~\ref{sec:cut-block-structure} analyzes cut-induced global gauge-sector ordering, and Sec.~\ref{sec:weight-response} presents the joint-distribution and flavor-pair responses together with the finite-chain and numerical controls.

\subsection{Benchmarks and numerical resolution}
\label{sec:benchmarks}

We first validate the Hamiltonian construction, ground-state calculations, and symmetry-resolved Schmidt spectra against exact diagonalization. For $N_s=2,3,4$, the MPO agrees with sparse ED at matrix level, while small-system DMRG reproduces the energies and resolved Schmidt weights. Flavor exchange provides an additional symmetry check. Table~\ref{tab:validation} summarizes these benchmarks. Solver details and convergence diagnostics are given in Appendix~\ref{app:validation}, while reciprocal checks are reported in Appendix~\ref{app:derived-logarithmic}.

\begin{table}[!htbp]
\caption{Numerical validation hierarchy. The listed errors are maxima or representative benchmark values for the stated comparison. Hamiltonian matrix elements and energy differences are reported in units of $g$, energy variances in units of $g^2$, and probability differences are dimensionless.}
\label{tab:validation}
\centering
\scriptsize
\begin{tabularx}{\textwidth}{@{}>{\raggedright\arraybackslash}p{0.21\textwidth} c >{\raggedright\arraybackslash}X >{\raggedright\arraybackslash}X@{}}
\toprule
Validation & Sizes & Diagnostic & Result \\
\midrule
MPO versus sparse ED & $N_s=2,3,4$ & maximum matrix-element difference ($g$) & $1.78\times10^{-15}$ \\
DMRG versus ED & $N_s=4$ & sector-ground-state energy difference ($g$) & $1.20\times10^{-14}$ \\
sector DMRG versus sector ED & $N_s=8$ & sector-ground-state energy difference ($g$) & $4.53\times10^{-12}$ \\
sector DMRG benchmark & $N_s=8$ & energy variance ($g^2$) & total $5.21\times10^{-11}$; per site $6.51\times10^{-12}$ \\
resolved Schmidt spectrum & $N_s=2,3,4$ & ED--MPS block labels and probabilities & max probability difference $7.62\times10^{-14}$; 0 label mismatches \\
\bottomrule
\end{tabularx}
\end{table}
\FloatBarrier

The states entering the main analysis satisfy the convergence criteria of Sec.~\ref{sec:dmrg}. Successive-bond-dimension comparisons show that $p_\pm$, $W_F^{(-1)}$, $\mathcal P_F^{(-1)}$, and the derived logarithmic quantities are stable over the available increases. Detailed componentwise convergence diagnostics are given in Appendix~\ref{app:validation}.

\FloatBarrier

\subsection{Cut-class alternation and global flux-sector reordering}
\label{sec:cut-block-structure}

\subsubsection{Alternating bond profiles}

The cut classification of Sec.~\ref{sec:subsystem-charges} has a direct empirical counterpart in these states. Finite-interval Schwinger calculations and open-boundary tensor-network studies can exhibit strong boundary and profile effects \cite{buyens2016,okuda2023}, while boundary-sensitive symmetry resolution is known more generally \cite{bonsignori2021}. In our data, bond-by-bond entanglement profiles additionally display a pronounced alternating component tied to the staggered cut class. In the central-neighbor calculations, the spread (maximum minus minimum) among the three cuts reached approximately 0.334 in the entropy and approximately 1.69 in the global entanglement-energy gap.

\begin{figure}[!htbp]
\centering
\includegraphics[width=\textwidth]{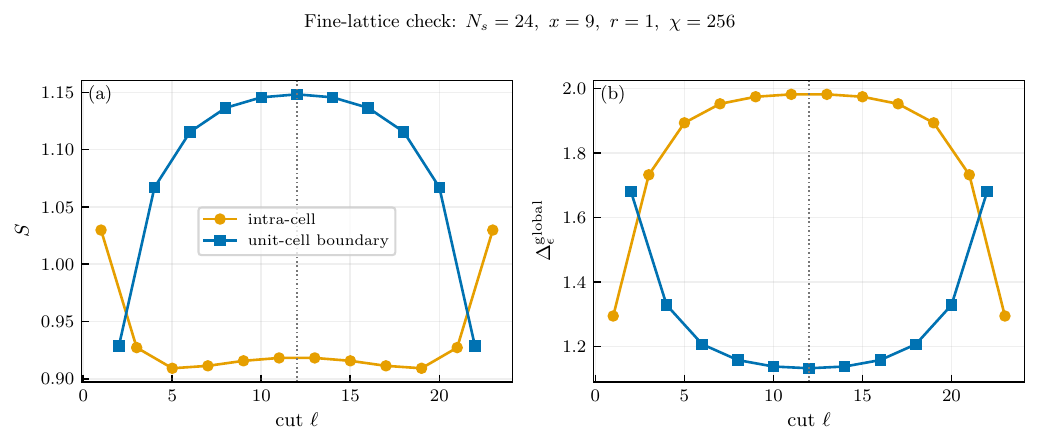}
\caption{Bond-resolved (a) entropy and (b) global entanglement-energy gap for the $N_s=24$, $x=9$, $r=1$, $\chi=256$ state. Intra-cell and unit-cell-boundary cuts are distinguished, and the dotted line marks the central unit-cell boundary $\ell=12$. Both quantities are dimensionless; entropy is reported in nats and $\ell$ is an integer cut index.}
\label{fig:cut-block-ordering}
\end{figure}

The bond profiles in Fig.~\ref{fig:cut-block-ordering} show the alternation beyond the central-neighbor controls. Class-resolved centered-cut values confirm the same ordering when either cut class occupies the geometric center (Appendix~\ref{app:validation}, Table~\ref{tab:cut-pilot}).

\FloatBarrier
\subsubsection{Reordering of the globally leading flux sectors}
\label{sec:global-reordering}

The unresolved spectrum is obtained by sorting all retained entanglement levels $\epsilon_\alpha$ without regard to symmetry. Its first spacing,
\begin{equation}
\Delta_\epsilon^{\mathrm{global}}=\epsilon_2-\epsilon_1,
\label{eq:global-gap}
\end{equation}
may therefore compare different $(Q_A,F_A)$ blocks. This inter-flux spacing is well defined, but it is not an internal spacing of either block. For the $L_{-1}=0$, $\theta=\pi$ convention, $Q_A=0,-1,+1$ correspond respectively to shifted cut fluxes $+1/2,-1/2,+3/2$.

\begin{figure}[!tbp]
\centering
\includegraphics[width=\textwidth]{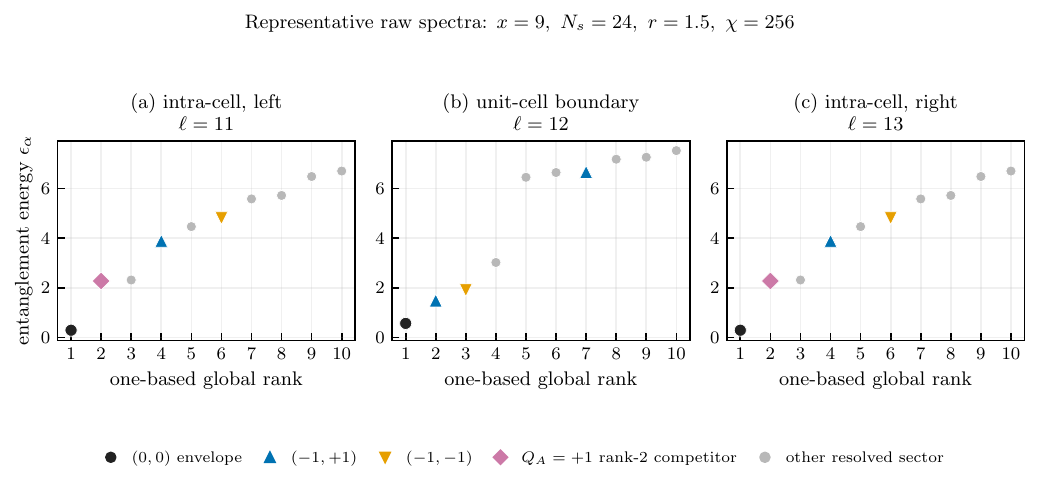}
\caption{Representative low-lying entanglement spectrum for the $x=9$, $N_s=24$, $r=1.5$, $\chi=256$ state at (a) the left adjacent intra-cell cut, (b) the selected unit-cell boundary, and (c) the right adjacent intra-cell cut. Each panel shows the first ten global levels on identical axes. The selected boundary has the leading $Q_A=0\leftrightarrow-1$ pair, whereas both adjacent cuts have $Q_A=0\leftrightarrow+1$; the tracked $Q_A=-1,F_A=\pm1$ levels remain identifiable at higher ranks. Entanglement energies are dimensionless, and global rank is an ordinal index.}
\label{fig:global-rank-cut}
\end{figure}

Throughout the complete dataset, the first two global levels at selected complete-unit-cell boundaries lie in $Q_A=0$ and $Q_A=-1$, whereas every reflection-related pair of adjacent intra-cell cuts gives $Q_A=0\leftrightarrow+1$. The $Q_A=-1,F_A=\pm1$ partner minima remain identifiable, moving from ranks 2 and 3 at the selected boundary to ranks 4 and 5--7 intra-cell. Global Schmidt-level sorting and fixed-sector flavor response therefore carry different information. Figure~\ref{fig:global-rank-cut} illustrates the resolved spectra for one representative state.

\subsection{Joint gauge--flavor Schmidt-probability response}
\label{sec:weight-response}

\subsubsection{Complementary gauge-charge pairs and the joint distribution}

The complete Schmidt distribution resolves a complementary pair in the $Q_A=+1$ sector as well as the tracked $Q_A=-1$ pair. Applying Eq.~\eqref{eq:pair-sum-polarization} to both $q=\pm1$ sectors, both partner blocks have nonzero probability at every analyzed cut, and the $F_A=\pm1$ pair exhausts the $Q_A=+1$ probability to numerical precision. The full gauge-charge marginals $p_{Q_A}(\pm1)$ obey the same cut-dependent ordering as the complementary-pair weights. The latter undergo a complete reversal:
\begin{equation}
0.00203\leq
\frac{W_F^{(+1)}}{W_F^{(-1)}}
\leq0.01111
\quad\text{at unit-cell boundaries},
\qquad
4.453\leq
\frac{W_F^{(+1)}}{W_F^{(-1)}}
\leq8.777
\quad\text{intra-cell}.
\label{eq:complementary-weight-switch}
\end{equation}
Thus the $Q_A=-1$ pair dominates at every selected unit boundary, whereas the $Q_A=+1$ pair dominates for every reflection-related intra-cell pair. Together with the global-rank identities in Sec.~\ref{sec:global-reordering}, this confirms the complementary pair switch. The corresponding conditional polarizations are not mirror copies: at unequal mass $\mathcal P_F^{(+1)}$ is negative at unit boundaries and positive intra-cell, while $\mathcal P_F^{(-1)}$ is positive at both cuts.

\begin{figure}[!tbp]
\centering
\includegraphics[width=\textwidth]{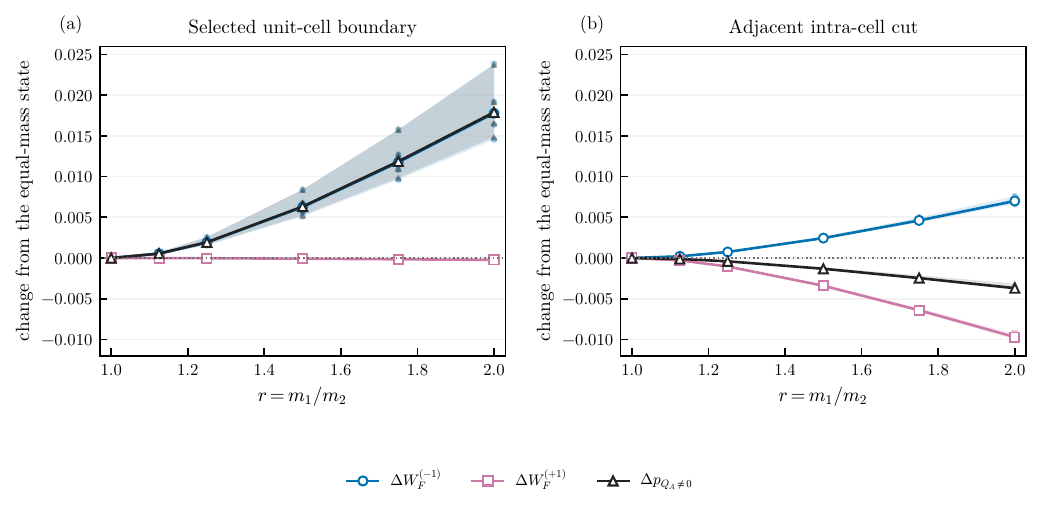}
\caption{Reference-subtracted reorganization of the joint Schmidt-probability distribution at $gL_{\mathrm{phys}}=8$. Panel (a) shows the selected unit-cell boundary and panel (b) the mirror-combined adjacent intra-cell cut class. The plotted quantities are the changes from the equal-mass state at the same $x$: the two complementary pair weights and the total non-neutral probability $p_{Q_A\ne0}=1-p_{Q_A=0}$. Lines show the pointwise median over $x=4,6.25,9,12.25$, shaded envelopes show their deterministic minimum--maximum range, and faint markers show all four values. The envelopes describe lattice-spacing dependence and are not statistical uncertainties. All plotted quantities, $x$, $r$, and $gL_{\mathrm{phys}}$ are dimensionless.}
\label{fig:joint-probability-distribution}
\end{figure}

Figure~\ref{fig:joint-probability-distribution} shows the changes in the major joint-probability components relative to the equal-mass state at fixed $gL_{\mathrm{phys}}=8$. At unit-cell boundaries, $\Delta W_F^{(-1)}$ closely tracks the positive change in total non-neutral probability, while the $Q_A=+1$ pair changes only weakly. At intra-cell cuts, $\Delta W_F^{(-1)}>0$ occurs together with a larger decrease of $W_F^{(+1)}$ and a net reduction of non-neutral probability. The former response is therefore non-neutral broadening dominated by the $Q_A=-1$ pair, whereas the latter is principally redistribution from the dominant $Q_A=+1$ pair toward $Q_A=-1$. These sign patterns hold at all four sampled spacings. The corresponding absolute joint distributions are shown in Appendix~\ref{app:validation}, Fig.~\ref{fig:absolute-joint-distribution}.

\subsubsection{Individual probabilities and pair weight}

The complete Schmidt vectors give $p_+$ and $p_-$ directly, without reconstructing either quantity from a truncated list of leading levels. At every sampled geometry and cut class, increasing $r$ gives
\begin{equation}
p_+\uparrow,
\qquad
p_-\downarrow,
\qquad
W_F^{(-1)}\uparrow,
\qquad
\mathcal P_F^{(-1)}\uparrow
\label{eq:sampled-weight-monotonicity}
\end{equation}
at the sampled ratios. The simultaneous growth of $W_F^{(-1)}$ and $\mathcal P_F^{(-1)}$ is important: flavor imbalance both makes the partner pair more probable and biases the conditional distribution toward $F_A=+1$. It is not merely moving a fixed amount of probability from $p_-$ to $p_+$.

Across the full six-ratio $gL_{\rm phys}=8$ scan, the selected-boundary pair weight is approximately $0.35$--$0.40$, whereas the mirror-combined intra-cell value is approximately $0.018$--$0.038$. Despite this smaller weight, the intra-cell polarization grows more rapidly and reaches $0.61$--$0.76$ at $r=2$ over the four spacings, compared with $0.32$--$0.42$ at the unit-cell boundary.

\subsubsection{Cut-class contrast}

\begin{figure}[!htbp]
\centering
\includegraphics[width=\textwidth]{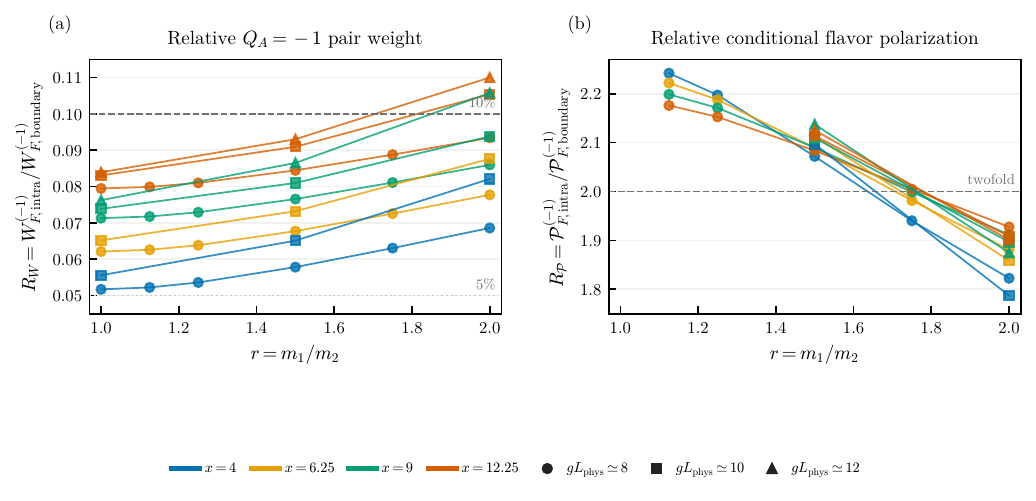}
\caption{State-matched intra-to-boundary ratios versus mass ratio. Panel (a) shows $R_W=W_{F,\mathrm{intra}}^{(-1)}/W_{F,\mathrm{boundary}}^{(-1)}$ for all 42 states, including equal mass. Panel (b) shows $R_{\mathcal P}=\mathcal P_{F,\mathrm{intra}}^{(-1)}/\mathcal P_{F,\mathrm{boundary}}^{(-1)}$ for the 32 unequal-mass states. At $r=1$, flavor symmetry gives $\mathcal P_{F,\mathrm{intra}}^{(-1)}=\mathcal P_{F,\mathrm{boundary}}^{(-1)}=0$, so $R_{\mathcal P}$ is undefined and no equal-mass point is plotted in panel (b). Color encodes $x$, marker shape encodes $gL_{\mathrm{phys}}\simeq8,10,12$, and thin lines connect only sampled ratios from the same finite-chain family as guides to the eye. All plotted quantities are dimensionless.}
\label{fig:matched-cut-probabilities}
\end{figure}

The two aspects of the response move in opposite directions under a change of cut. For each unequal-mass state we compare the selected unit-cell boundary with the arithmetic mean of its reflection-related intra-cell block weights. The resulting ratios are
\begin{equation}
0.0523\leq
\frac{W_{F,\mathrm{intra}}^{(-1)}}{W_{F,\mathrm{boundary}}^{(-1)}}
\leq0.1100,
\qquad
1.787\leq
\frac{\mathcal P_{F,\mathrm{intra}}^{(-1)}}{\mathcal P_{F,\mathrm{boundary}}^{(-1)}}
\leq2.242.
\label{eq:matched-cut-ranges}
\end{equation}
Thus the intra-cell cut retains only about $5\%$--$11\%$ of the partner-pair probability found at the unit boundary, but has the larger conditional flavor polarization. The polarization ratio is not constant: over the complete six-ratio families it decreases systematically from $2.176$--$2.242$ at $r=1.125$ to $1.823$--$1.928$ at $r=2$. Figure~\ref{fig:matched-cut-probabilities} plots the state-matched ratios against $r$. At every sampled geometry, the weight ratio increases over the available mass ratios, whereas the polarization ratio decreases over the available unequal-mass ratios. The weight suppression and polarization enhancement are therefore parameter dependent rather than constant cut factors.

The increase of $R_W$ reflects larger fractional, not absolute, growth of the smaller intra-cell pair, whereas the decrease of $R_{\mathcal P}$ largely reflects nonlinear compression by the bounded polarization map. These trends are interpreted in Sec.~\ref{sec:discussion}.

\FloatBarrier

The block-minimum difference $D_F$ remains a valid full-spectrum entanglement-energy difference, but Eq.~\eqref{eq:df-derived-decomposition} now fixes its role. Across the same 32 unequal-mass states and both cut classes, the normalized-block-minimum correction never exceeds $1.292\%$ of $|D_F|$. The logarithmic response and its cut contrast are therefore almost entirely carried by the pair-normalized conditional flavor asymmetry at fixed $Q_A=-1$; the complete decomposition and fit diagnostics are reported in Appendix~\ref{app:derived-logarithmic}.

Complete-spectrum reconstruction and successive-$\chi$ comparisons show that $W_F^{(\pm1)}$ and $\mathcal P_F^{(\pm1)}$ are stable over the available bond dimensions, while reflection-related intra-cell cuts agree to numerical precision. An exactly centered control calculation reproduces $W_{F,\mathrm{intra}}^{(-1)}\ll W_{F,\mathrm{unit}}^{(-1)}$ and $\mathcal P_{F,\mathrm{intra}}^{(-1)}>\mathcal P_{F,\mathrm{unit}}^{(-1)}$, confirming that the contrast follows cut class rather than the one-site centering offset. Appendix~\ref{app:validation} reports the full truncation, convergence, centering, and finite-OBC controls.

\section{Discussion}
\label{sec:discussion}

The block weights $p_{q,f}$ are the joint probability distribution of subsystem gauge charge and flavor imbalance for the chosen matter-site bipartition. Marginalizing over $f$ gives the gauge-charge organization $p_{Q_A}(q)$, whereas conditioning at fixed $q$ gives $p_{F_A\mid Q_A}(f\mid q)$. The two $Q_A=\pm1,F_A=\pm1$ pairs demonstrate that cut geometry reorganizes this distribution rather than merely suppressing one preselected sector: $Q_A=-1$ dominates at unit boundaries and $Q_A=+1$ intra-cell. For either $q=\pm1$, $W_F^{(q)}$ is the absolute joint weight of the $F_A=\pm1$ pair, while $\mathcal P_F^{(q)}$ is the normalized contrast of those two conditional flavor probabilities. Their different cut dependence remains important: a small pair weight does not imply weak conditional sensitivity, and a large pair weight does not make the pair maximally polarized.

The increase of $R_W$ means that $W_{F,\mathrm{intra}}^{(-1)}$ grows by a larger fraction, not by a larger absolute amount, than $W_{F,\mathrm{boundary}}^{(-1)}$. At unit-cell boundaries the $Q_A=-1$ pair already dominates the non-neutral probability and its growth accompanies non-neutral broadening. At intra-cell cuts the initially smaller $Q_A=-1$ pair grows while the dominant $Q_A=+1$ pair and the total non-neutral probability decrease, consistent with a redistribution principally from $Q_A=+1$ toward $Q_A=-1$.

At equal mass the continuum theory has an $SU(2)/\mathbb Z_2\simeq SO(3)$ flavor symmetry. Writing the real mass matrix as $\widetilde M=m\mathbb I_2+\delta m\,\sigma^3$ gives
\begin{equation}
m=m_{\rm geom}\cosh\!\left(\frac{\eta}{2}\right),
\qquad
\delta m=m_{\rm geom}\sinh\!\left(\frac{\eta}{2}\right),
\qquad
\eta=\ln r,
\end{equation}
so flavor-invariant and flavor-odd responses are respectively even and odd in $\eta$ \cite{hetrick1995,smilga1997,dempsey2024,albandea2025,cuomo2026}. The present data resolve that odd response in subsystem probabilities. The nearly power-law behavior of $p_-/p_+$ over $1\le r\le2$ has a small but resolved scale-dependent effective exponent; because it depends on cut, spacing, and finite length, it is retained as an Appendix parameterization rather than a universal exponent.

For the polarization ratio, define $u_c=-D_{F,c}^{\mathrm{weight}}/2=\operatorname{artanh}\mathcal P_{F,c}^{(-1)}>0$ for $c=\mathrm{boundary},\mathrm{intra}$ and $r>1$. Then $R_{\mathcal P}=\tanh u_{\mathrm{intra}}/\tanh u_{\mathrm{boundary}}$. For smooth nonzero linear responses, its one-sided equal-mass limit is the ratio of the corresponding linear coefficients. Because $u_{\mathrm{intra}}>u_{\mathrm{boundary}}$ and $\tanh u/u$ decreases with $u>0$, the bounded map compresses $R_{\mathcal P}$ below the underlying log-odds ratio as the imbalance grows. The log-odds ratio itself also decreases slightly over the sampled points, so the observed decline of $R_{\mathcal P}$ is largely, but not exclusively, a consequence of this nonlinear compression.

The cut class organizes more than one layer of the reduced state. It determines which non-neutral gauge pair carries most probability and competes globally, while mass imbalance redistributes probability both between charge sectors and between flavors within a pair.  The $Q_A=+1$ polarization changes sign between the two cut classes, so there is no cut-independent polarization law shared by the complementary pairs. Consequently, an unresolved global entanglement gap cannot distinguish a change of gauge-sector identity from a change in fixed-sector probabilities. The derived $D_F$ is useful as a logarithmic transform of $\mathcal P_F^{(-1)}$ up to a small normalized-block correction, but the joint distribution exposes the underlying probability reorganization directly.

The atomic-reference analysis identifies the lowest nonzero sector-selection processes; quantitative probabilities are determined from the MPS results. The conclusions refer to the finite open-chain, Gauss-law-eliminated matter-site bipartition studied here. Continuum, thermodynamic, and alternative boundary or factorization prescriptions remain beyond the present analysis.

\section{Summary}
\label{sec:summary}

The complete joint Schmidt-probability distribution reveals a cut-dependent switch of the dominant non-neutral gauge-charge pair: $(Q_A,F_A)=(-1,\pm1)$ dominates at selected unit-cell boundaries, whereas the complementary $(+1,\pm1)$ pair dominates at adjacent intra-cell cuts. The full gauge-charge marginals $p_{Q_A}(\pm1)$ obey the same hierarchy. At unequal mass, $\mathcal P_F^{(+1)}$ is negative at unit-cell boundaries and positive intra-cell, whereas $\mathcal P_F^{(-1)}$ is positive for both cut classes; the conditional flavor response is therefore both charge and cut dependent. Within the tracked $Q_A=-1$ pair, increasing the mass ratio under the fixed flavor ordering raises $p_+$, lowers $p_-$, and increases both the combined weight $W_F^{(-1)}$ and the conditional polarization $\mathcal P_F^{(-1)}$ throughout the sampled geometries. For the matched unequal-mass states, the cut ratios satisfy $0.0523\leq R_W\leq0.1100$, with $R_W=W_{F,\mathrm{intra}}^{(-1)}/W_{F,\mathrm{boundary}}^{(-1)}$, and $1.787\leq R_{\mathcal P}\leq2.242$, with $R_{\mathcal P}=\mathcal P_{F,\mathrm{intra}}^{(-1)}/\mathcal P_{F,\mathrm{boundary}}^{(-1)}$. The weight ratio increases along every sampled mass-ratio sequence, whereas the polarization ratio decreases along the unequal-mass sequences; neither contrast is a parameter-independent cut factor.

The increase of $W_F^{(-1)}$ nearly accounts for the positive change in total non-neutral probability at unit-cell boundaries, identifying an overall broadening of the charge distribution. Intra-cell, $W_F^{(-1)}$ increases while the dominant $Q_A=+1$ pair and the total non-neutral probability both decrease, identifying redistribution among non-neutral sectors rather than uniform broadening. The full-spectrum difference satisfies $D_F=D_F^{\mathrm{weight}}+D_F^{\mathrm{conf}}$, where $D_F^{\mathrm{weight}}=\ln(p_-/p_+)=-2\operatorname{artanh}\mathcal P_F^{(-1)}$ and the normalized-block correction $D_F^{\mathrm{conf}}$ never exceeds $1.292\%$ of $|D_F|$ over the sampled unequal-mass data. Consequently, $D_F$ is predominantly a logarithmic diagnostic of the pair-normalized conditional flavor asymmetry at fixed $Q_A=-1$. Together with the global-rank switch, these results separate gauge-charge ordering, complementary-pair weight, and conditional flavor polarization.

\data{The complete derived datasets and analysis scripts will be deposited in a public repository upon submission. They can be made available to reviewers during peer review and are available from the corresponding author upon reasonable request.}

\appendix

\section*{APPENDIX}

\section{Exact lattice construction and subsystem-charge bookkeeping}

\subsection{Exact electric-term MPO}\label{app:electric-mpo}

The eliminated-gauge-field electric energy contains cumulative charge squares.
Let \(\kappa=g^2a/2\), \(b_\theta=L_{-1}+\theta/(2\pi)\), and \(\delta_R=0\) for the internal-link convention. Define

\begin{equation}
\left(b_\theta+\sum_{k\le n}Q_k\right)^2
=
b_\theta^2
+2b_\theta\sum_{k\le n}Q_k
+\sum_{k\le n}Q_k^2
+2\sum_{i<j\le n}Q_iQ_j.
\end{equation}

After summing this expression over the internal links, define

\begin{equation}
R=N_s-1+\delta_R,
\qquad
w_i=\max(R-i,0).
\end{equation}

Then

\begin{equation}
H_E=E_{\mathrm{const}}+\sum_i A_iQ_i+\sum_i B_iQ_i^2+\sum_{i<j}V_{ij}Q_iQ_j,
\end{equation}

with

\begin{equation}
E_{\mathrm{const}}=\kappa R b_\theta^2,
\quad A_i=2\kappa b_\theta w_i,
\quad B_i=\kappa w_i,
\quad V_{ij}=2\kappa\max[R-\max(i,j),0].
\end{equation}

For \(\delta_R=0\), \(w_{N_s-1}=0\), so the rightmost matter-site charge does not enter the internal-link electric energy. The exact three-state electric MPO uses

\begin{equation}
W_i=
\begin{pmatrix}
\mathbf 1 & Q_i & e_i\\
0 & \mathbf 1 & V_iQ_i\\
0 & 0 & \mathbf 1
\end{pmatrix},
\end{equation}

where \(i=0,\ldots,N_s-1\), \(V_i=2\kappa w_i\), and \(e_i=A_iQ_i+B_iQ_i^2\). With the displayed upper-triangular orientation, the boundary vectors are

\begin{equation}
v_L=(1,0,0),
\qquad
v_R=(0,0,1)^{\mathsf T}.
\end{equation}

The operator-independent contribution \(E_{\mathrm{const}}\mathbf 1\) is added once to \(e_0\), matching the MPO construction rather than being represented by a second identity MPO. This yields bond dimension 3 independently of \(N_s\). Including the hopping and mass terms gives full-Hamiltonian bond dimension 7 for the conventions used here.

\subsection{Subsystem-charge mapping}
\label{app:charge-map}

For a cut after \(\ell\) physical sites,

\begin{equation}
Q_A=N_A-2\lfloor\ell/2\rfloor,
\qquad F_A=N_{1,A}-N_{2,A}.
\end{equation}

The tensor-network representation labels each block by \((N_A,F_A^{\mathrm{backend}})\), with $F_A=F_A^{\mathrm{backend}}$, and the analysis maps it to the physical \((Q_A,F_A)\). Sector-blocked ED singular-value decompositions provide the reference mapping. Degenerate singular subspaces are never assigned charges from an unresolved SVD.

\subsection{Spatial-coordinate convention}
\label{app:coordinate-convention}

The matter sites lie at $z_n=(n+\frac{1}{2})a$ between the geometric boundaries $z=0$ and $z=N_sa$. Consequently $L_{\mathrm{phys}}=N_sa$. The endpoint-to-endpoint distance between the outermost matter sites, $(N_s-1)a$, is a different geometric quantity and is not used as the volume label.

\section{Numerical validation, auxiliary diagnostics, and exact symmetry checks}
\label{app:validation}

\subsection{Cut-class entropy and global-gap control}

The class-resolved central-neighbor values in Table~\ref{tab:cut-pilot} show that unit-cell boundaries have the larger entropy and smaller global gap whether that cut class occurs at the geometric center or at its two neighbors.

\begin{table}[!tbp]
\caption{Cut-class-resolved entropy and global entanglement-energy gap in the \(r=1\) and \(r=2\) central-neighbor calculations. All rows use \(x=4\) and \(\chi=256\). When a class is represented by the two symmetric neighboring cuts, their mean is shown; their maximum left--right difference over both observables is \(3.3\times10^{-10}\). The entropy uses the natural logarithm and is reported in nats; the global gap is a difference of dimensionless entanglement energies and is dimensionless.}
\label{tab:cut-pilot}
\centering
\scriptsize
\begin{tabular}{@{}c c c c c c@{}}
\toprule
\(N_s\) & \(r\) & \(S_{\mathrm{intra}}\) & \(S_{\mathrm{unit}}\) & \(\Delta_{\epsilon,\mathrm{intra}}^{\mathrm{global}}\) & \(\Delta_{\epsilon,\mathrm{unit}}^{\mathrm{global}}\) \\
\midrule
18 & 1 & 0.733201 & 1.064565 & 2.323677 & 1.230324 \\
18 & 2 & 0.745423 & 1.065650 & 2.349588 & 0.748087 \\
24 & 1 & 0.738580 & 1.072955 & 2.329807 & 1.213897 \\
24 & 2 & 0.770881 & 1.087687 & 2.341787 & 0.647335 \\
\bottomrule
\end{tabular}
\end{table}

\begin{figure}[!tbp]
\centering
\includegraphics[width=\textwidth]{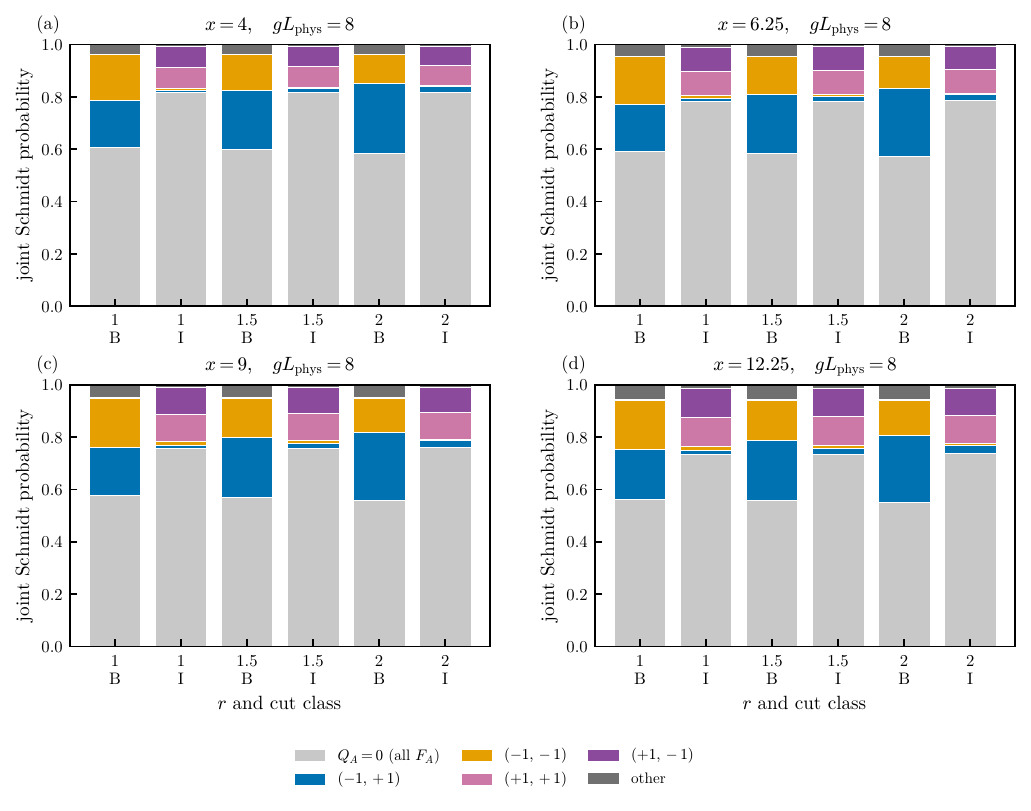}
\caption{Absolute major-block composition of the joint Schmidt-probability distribution at $gL_{\mathrm{phys}}=8$ for $r=1,1.5,2$. Each mass ratio is shown at the selected unit-cell boundary (B) and the mirror-combined intra-cell cut class (I) for all four sampled spacings. The six categories are all $Q_A=0$ blocks, the two $Q_A=-1,F_A=\pm1$ partners, the two $Q_A=+1,F_A=\pm1$ partners, and all remaining sectors. Every bar sums to unity using the complete Schmidt spectrum. All probabilities and the parameters $x$, $r$, and $gL_{\mathrm{phys}}$ are dimensionless.}
\label{fig:absolute-joint-distribution}
\end{figure}

Figure~\ref{fig:absolute-joint-distribution} records the absolute joint-probability composition underlying the reference-subtracted response in Fig.~\ref{fig:joint-probability-distribution}.

\subsection{ED basis maps and solver prescriptions}

A basis state is represented by a $2N_s$-mode occupation bit string in the site-major ordering of Sec.~\ref{sec:local-basis}. Fermionic actions and Jordan--Wigner signs are applied directly to the bit strings, and full or symmetry-restricted Hamiltonians are assembled as complex sparse matrices. In the MPS representation, local fermion operators carry the corresponding metadata so that inter-site parity strings are inserted consistently; small-system matrix reconstruction tests verify the resulting signs.

The matrix-level validation compares hopping, mass, electric, and full MPOs with ED at $N_s=2,3,4$. DMRG energies, term energies, local charges, electric-flux moments, Schmidt probabilities, and subsystem labels are compared with ED at small size. A dominant-block assignment is marked ambiguous when the difference between the two largest block weights is below $10^{-6}$; no point in the main analysis is ambiguous. The unrestricted ED basis follows the site-major bit-string order, and symmetry-restricted bases retain explicit forward and inverse maps. Dense Hermitian diagonalization is used when the matrix dimension is at most 16 or the requested eigenpairs span the matrix. If sparse diagonalization does not return the requested eigenpairs, dense Hermitian diagonalization is used for matrices of dimension at most 256. Solver residuals were monitored in all validation calculations.

\subsection{Convergence and auxiliary bulk-window diagnostics}\label{app:bulk-window}

The default comparison uses $\chi=128,256$, and two matched $r=1,1.5,2$ triplets were extended to $\chi=384$. At $x=12.25,N_s=40$, the $\chi=128$ to 256 total-energy drifts are $1.15$--$1.44\times10^{-6}$; at $x=12.25,N_s=36$, the equal-mass reference has drift $1.080\times10^{-6}$. These cases are evaluated at $\chi=384$, where they satisfy the stated criteria. A reference-subtracted response is retained only when both the unequal-mass state and its matched reference pass the convergence tests; the remaining matched sets pass at $\chi=256$.

For the 32 unequal-mass states, Table~\ref{tab:component-convergence} summarizes direct componentwise changes between the two largest available bond dimensions. The mirror-related intra-cell values are combined before differencing. The finite-MPS states were generated with the specified TeNPy truncation controls, including $\mathrm{svd\_min}=10^{-12}$. For the block-weight reconstruction, every Schmidt value present in each available MPS state was summed without the $10^{-12}$ probability filter used for low-lying level tables.

\begin{table}[!htbp]
\caption{Componentwise convergence from complete Schmidt spectra. Entries are the median and maximum absolute successive-$\chi$ changes over 32 unequal-mass states. The two cut classes are summarized independently. All entries are dimensionless.}
\label{tab:component-convergence}
\centering
\scriptsize
\begin{tabular}{@{}l c c c c@{}}
\toprule
cut class & quantity & median & maximum & $N$ \\
\midrule
unit-cell boundary & $|\Delta p_+|$ & $1.25\times10^{-8}$ & $2.25\times10^{-7}$ & 32 \\
 & $|\Delta p_-|$ & $6.22\times10^{-9}$ & $1.40\times10^{-7}$ & 32 \\
 & $|\Delta W_F^{(-1)}|$ & $1.88\times10^{-8}$ & $3.26\times10^{-7}$ & 32 \\
 & $|\Delta\mathcal P_F^{(-1)}|$ & $3.53\times10^{-9}$ & $1.34\times10^{-7}$ & 32 \\
 & $|\Delta D_F^{\mathrm{weight}}|$ & $8.04\times10^{-9}$ & $3.17\times10^{-7}$ & 32 \\
 & $|\Delta D_F^{\mathrm{conf}}|$ & $3.62\times10^{-9}$ & $1.38\times10^{-7}$ & 32 \\
\addlinespace
mirror-combined intra-cell & $|\Delta p_+|$ & $1.24\times10^{-8}$ & $2.36\times10^{-7}$ & 32 \\
 & $|\Delta p_-|$ & $6.82\times10^{-9}$ & $1.81\times10^{-7}$ & 32 \\
 & $|\Delta W_F^{(-1)}|$ & $1.87\times10^{-8}$ & $4.04\times10^{-7}$ & 32 \\
 & $|\Delta\mathcal P_F^{(-1)}|$ & $2.00\times10^{-7}$ & $2.87\times10^{-6}$ & 32 \\
 & $|\Delta D_F^{\mathrm{weight}}|$ & $5.16\times10^{-7}$ & $7.00\times10^{-6}$ & 32 \\
 & $|\Delta D_F^{\mathrm{conf}}|$ & $1.49\times10^{-7}$ & $2.23\times10^{-6}$ & 32 \\
\bottomrule
\end{tabular}
\end{table}

For $M=N_s-1$ internal links, the default bulk window is the inclusive-exclusive interval $[j_0,j_1)$. Define
\begin{equation}
\begin{aligned}
w_0&=\max\!\left[2,\operatorname{round}_{\mathrm{ties\mbox{-}to\mbox{-}even}}\!\left(\frac{M}{2}\right)\right],\\
\widetilde w&=\begin{cases}
w_0,&w_0\bmod2=M\bmod2,\\
w_0+1,&\text{otherwise},
\end{cases}\\
w&=\min(M,\widetilde w),\qquad j_0=(M-w)/2,\qquad j_1=j_0+w.
\end{aligned}
\end{equation}
The cap covers the small validation systems; for $N_s=2$ it gives $M=1$ and $[j_0,j_1)=[0,1)$. Let $[s_0,s_1)$ be the deterministic central 50\% site window obtained from the same rounding, parity-matching, and profile-length-cap prescription, with the site-profile length $N_s$ in place of the internal-link length $M$. The two monitored bare staggered-occupation diagnostics are
\begin{equation}
C_{\pm}^{\mathrm{bulk}}
=
\frac{1}{s_1-s_0}
\sum_{n=s_0}^{s_1-1}
(-1)^n
\left(
\langle n_{n,1}\rangle
\pm
\langle n_{n,2}\rangle
\right).
\end{equation}
where $C_+^{\mathrm{bulk}}$ and $C_-^{\mathrm{bulk}}$ are called the condensate sum and condensate difference, respectively. No additional staggered-background subtraction or continuum renormalization is applied to these convergence diagnostics. Here $L_{\mathrm{bulk}}$ denotes the arithmetic mean of the bare link flux over the deterministic central internal-link window. With $\Delta$ denoting the change between the two largest available bond dimensions, we define
\begin{equation}
\Delta O_{\mathrm{bulk}}
=
\max\!\left(
|\Delta L_{\mathrm{bulk}}|,
|\Delta C_+^{\mathrm{bulk}}|,
|\Delta C_-^{\mathrm{bulk}}|
\right).
\end{equation}
For a fractional interval $[f_0,f_1]$, the integer window is $[\lceil f_0M\rceil,\lceil f_1M\rceil)$. The alternatives $[0.35,0.65]$ and $[0.40,0.60]$ are used only for bulk-profile calculations with $N_s\ge16$. Small-system ED--MPS validation instead compares link-resolved moments directly. The bare-link-flux, condensate, and R\'enyi-2 observables enter only the convergence monitor $\Delta O_{\rm bulk}$ and boundary-sensitivity checks; they are not interpreted as physical results in the present work.

Two $x=12.25$ systems contain the three-point ladder $\chi=128,256,384$ for both the equal-mass reference and $r=2$. The successive changes in $D_F$ are given below, together with the relative change for all selected unit-cell-boundary states with unequal masses.

\begin{table}[!ht]
\caption{Three-point bond-dimension convergence sequences for $D_F$ at $r=2$. The observable and its successive differences are dimensionless; $gL_{\mathrm{phys}}$, $r$, and $\chi$ are dimensionless, while $N_s$ is a site count.}
\label{tab:chi-ladders}
\centering
\small
\begin{tabular}{@{}l c c@{}}
\toprule
system & $|D_{F,256}-D_{F,128}|$ & $|D_{F,384}-D_{F,256}|$ \\
\midrule
$gL_{\rm phys}=10.29$, $N_s=36$ & $4.47\times10^{-7}$ & $1.07\times10^{-9}$ \\
$gL_{\rm phys}=11.43$, $N_s=40$ & $8.57\times10^{-7}$ & $1.78\times10^{-9}$ \\
\bottomrule
\end{tabular}
\end{table}

\FloatBarrier

Across all analyzed states, the largest combined probability of the two target blocks below the low-level tabulation threshold $\lambda=10^{-12}$ is $9.97\times10^{-12}$, at most $3.15\times10^{-10}$ of their combined weight. These tail levels are included in the reported $p_\pm$.

As a separate finite-OBC control, we fit the absolute connected covariance of the bare electric field over the central 50\% of links to an exponential and denote the resulting operational scale by $\xi_E^{\mathrm{op}}$. Across the 42 analyzed states, $\xi_E^{\mathrm{op}}=1.523$--$4.184$ lattice sites and $N_s/\xi_E^{\mathrm{op}}=7.789$--$14.454$; the median and maximum successive-$\chi$ relative changes are $3.05\times10^{-4}$ and $3.48\times10^{-3}$. All fits satisfy the stated quality criteria at both available bond dimensions. This quantity is not a transfer-matrix correlation length, contains no Schmidt-cut input, and is not used as evidence for thermodynamic convergence.

\subsection{Cut-selection, centering, and finite-length controls}
\label{app:cut-classification}

With the natural cell \((2j,2j+1)\), a cut \(\ell\) lies between sites \(\ell-1\) and \(\ell\). Odd \(\ell\) is an intra-cell even-to-odd cut; even \(\ell\) is an odd-to-even unit-cell-boundary cut. Consequently, the central cut is a unit-cell boundary for \(N_s\bmod4=0\) and an intra-cell cut for \(N_s\bmod4=2\). The main entanglement observable uses the unit-cell boundary minimizing \(|\ell-N_s/2|\). If the two sides are equidistant, the left boundary (the smaller \(\ell\)) is selected deterministically; the reflection-related partner is retained as a consistency check.

The existing $N_s=18,x=4$ states reverse the centering pattern of the main $N_s\bmod4=0$ grid: cut $\ell=9$ is an exactly centered intra-cell cut and $\ell=8,10$ are the adjacent unit boundaries.  Table~\ref{tab:probability-centering-control} reports the complete-Schmidt probabilities.  At $r=2$ the centered intra-cell cut still has much smaller $W_F^{(-1)}$ and larger $\mathcal P_F^{(-1)}$ than either adjacent unit boundary.  The mean of the two $N_s=18$ unit boundaries also lies between the central-unit values at $N_s=16$ and 20 for $p_\pm$, $W_F^{(-1)}$, and $\mathcal P_F^{(-1)}$.  This isolates cut class from the one-site centering offset.

\begin{table}[!ht]
\caption{Reversed-centering control at $x=4$, $N_s=18$, and $\chi=256$, evaluated from complete Schmidt spectra.  The geometric center is the intra-cell cut $\ell=9$; the unit-boundary entries are the mean of $\ell=8$ and 10, whose maximum difference in the displayed quantities is $3.8\times10^{-11}$.  All probabilities and polarizations are dimensionless.}
\label{tab:probability-centering-control}
\centering
\small
\begin{tabular}{@{}c l r r r r@{}}
\toprule
$r$ & cut class & $p_{-1,+1}$ & $p_{-1,-1}$ & $W_F^{(-1)}$ & $\mathcal P_F^{(-1)}$ \\
\midrule
1 & central intra-cell & $0.00967440$ & $0.00967440$ & $0.0193488$ & $-3.25\times10^{-11}$ \\
1 & adjacent unit mean & $0.17737415$ & $0.17737415$ & $0.3547483$ & $3.95\times10^{-12}$ \\
2 & central intra-cell & $0.02640104$ & $0.00312267$ & $0.0295237$ & $0.788464$ \\
2 & adjacent unit mean & $0.27435205$ & $0.10861952$ & $0.3829716$ & $0.432754$ \\
\bottomrule
\end{tabular}
\end{table}

\FloatBarrier

At fixed $x$, fixed $r$, and fixed cut class, the available coverage contains two or three lengths rather than a uniform Cartesian grid.  Across all 24 such sequences, $W_F^{(-1)}$ is nondecreasing and $W_F^{(+1)}$ decreases.  For every unequal-mass sequence, the signed values of both conditional polarizations are nondecreasing along the available lengths; because $\mathcal P_F^{(+1)}$ is negative at unit-cell boundaries, this statement does not imply a common trend in its magnitude across cut classes.  These are finite-length sensitivities, not length-scaling laws; in particular, the total non-neutral probability is not strictly monotone in every intra-cell sequence.

\FloatBarrier

\section{Atomic-reference sector-selection rules}
\label{app:atomic-hopping}

This section records only the lowest nonzero local processes and the sectors they generate. It is a formal hopping-order classification for the lattice Hamiltonian, not a continuum strong-coupling expansion and not an explanation of the measured Schmidt probabilities. Write
\begin{equation}
H_0=H_{\mathrm{mass}}+H_E,
\qquad
V=H_{\mathrm{hop}},
\qquad
t=\frac{1}{2a},
\qquad
\mu_\alpha=m_\alpha-\frac{g^2a}{4}>0.
\label{eq:atomic-split}
\end{equation}
The atomic reference configuration has every even site empty and both flavors on every odd site. For $L_{-1}=0$ and $\theta=\pi$, the shifted field $L_n+\theta/(2\pi)$ is $+1/2$ on every internal link. Moving flavor $\alpha$ from an occupied odd site to a neighboring empty even site costs
\begin{equation}
\Delta_\alpha=2\mu_\alpha=2m_\alpha-\frac{g^2a}{2}.
\label{eq:atomic-denominator}
\end{equation}
The shifted field on that link changes from $+1/2$ to $-1/2$, so the elementary dipole has no electric-energy cost at $\theta=\pi$.

\subsection{Unit-cell boundary}

Denote a doubly occupied site by $12$, an empty site by $0$, and the flavor other than $\alpha$ by $\bar\alpha$. At an even cut $\ell$, the first-order cross-cut process is
\begin{equation}
\left\lvert12\,\middle|\,0\right\rangle
\xrightarrow{\;c_{\ell,\alpha}^{\dagger}c_{\ell-1,\alpha}\;}
\left\lvert\bar\alpha\,\middle|\,\alpha\right\rangle .
\label{eq:unit-atomic-sequence}
\end{equation}
It transfers one fermion out of $A$ and therefore produces $Q_A=-1$. Since $F_A=N_{1,A}-N_{2,A}$,
\begin{equation}
\alpha=1:\ (Q_A,F_A)=(-1,-1),
\qquad
\alpha=2:\ (Q_A,F_A)=(-1,+1).
\label{eq:atomic-flavor-map}
\end{equation}
The only conclusion needed below is that the target sectors first occur after one hop at this cut.  No Schmidt-sector probability is inferred from that order count.

\subsection{Intra-cell cut and the absence of first- and second-order target paths}

At an odd cut the adjacent atomic occupation is instead $\lvert0\mid12\rangle$. The only allowed first-order cross-cut hop points into $A$,
\begin{equation}
\left\lvert0\,\middle|\,12\right\rangle
\xrightarrow{\;c_{\ell-1,\alpha}^{\dagger}c_{\ell,\alpha}\;}
\left\lvert\alpha\,\middle|\,\bar\alpha\right\rangle,
\label{eq:intra-first-order-plus}
\end{equation}
and produces $Q_A=+1$. Its shifted cut field changes from $+1/2$ to $+3/2$, giving the excitation energy
\begin{equation}
\Delta_\alpha^{(Q_A=+1)}=2\mu_\alpha+g^2a=2m_\alpha+\frac{g^2a}{2}.
\label{eq:intra-plus-denominator}
\end{equation}

Producing $Q_A=-1$ requires a hop from $A$ to $\bar A$. In the atomic state this process is blocked at both ends: its even-site source in $A$ is empty and its odd-site destination in $\bar A$ already contains both flavors. One preparatory hop can either fill the source or create a destination vacancy, but cannot do both. Hence no two-hop sequence reaches $Q_A=-1$. If the first hop instead crosses the cut in the allowed direction, it produces $Q_A=+1$ and one further hop can at most return to $Q_A=0$.

The shortest local target path contains three hops. For the four sites $(\ell-2,\ell-1\mid\ell,\ell+1)$, one ordering is
\begin{equation}
\begin{aligned}
\left\lvert12,0\,\middle|\,12,0\right\rangle
&\longrightarrow
\left\lvert\bar\alpha,\alpha\,\middle|\,12,0\right\rangle\\
&\longrightarrow
\left\lvert\bar\alpha,\alpha\,\middle|\,\bar\alpha,\alpha\right\rangle\\
&\longrightarrow
\left\lvert\bar\alpha,0\,\middle|\,12,\alpha\right\rangle .
\end{aligned}
\label{eq:intra-third-order-sequence}
\end{equation}
The first preparatory hop fills the source, the second creates the destination vacancy, and the third crosses the cut. The preparatory hops may be exchanged, but must involve the same flavor. The final configuration has lost one flavor-$\alpha$ fermion from $A$, and therefore has the labels in Eq.~\eqref{eq:atomic-flavor-map}. This proves that the target pair first appears at third order for the local atomic reference, while the complementary $Q_A=+1$ sector already appears at first order.

Nothing quantitative about the full Schmidt-sector weights follows from these orders at the simulated parameters.  In particular, $t/\Delta_2\simeq2.19$--$3.10$ at representative $r=2$ points, so the hopping expansion is not controlled.  Each sector probability sums all Schmidt values and all left--right configurations, including connected dressing, path interference, normalization, and higher-order processes.  The robust content of this appendix is therefore limited to the lowest nonzero order and sector identity; it is not used to explain the observed weight hierarchy, polarization strength, or cut ratios.

\section{Derived logarithmic diagnostics}
\label{app:derived-logarithmic}

\subsection{Reciprocal checks}\label{app:fit-diagnostics}

Flavor exchange was checked at $r=2\leftrightarrow0.5$ and $r=1.0625\leftrightarrow\frac{1}{1.0625}$. The near-equal calculations use $(x,N_s,gL_{\rm phys},\chi,\ell_\star)=(9,24,8.00,256,12)$ and $(12.25,28,8.00,256,14)$. Flavor-invariant spectra agree while $F_A$ labels exchange; the specified consistency tolerance is $10^{-8}$, and the largest reported reciprocal residual is below $2.5\times10^{-9}$. This tolerance is not a DMRG convergence threshold. The near-equal points test the expected odd response without defining a precision susceptibility.

\subsection{Odd-response parameterization over the six-ratio scan}\label{app:odd-response-fits}

Flavor exchange makes $D_F(\eta)$ odd, so a smooth response has the form
\begin{equation}
D_F(\eta)=c_1\eta+c_3\eta^3+\mathcal O(\eta^5).
\label{eq:df-odd-expansion}
\end{equation}
Table~\ref{tab:df-odd-fits} gives the eight no-intercept fits to this symmetry-constrained expansion. ``Intra'' denotes the arithmetic mean of the two reflection-related adjacent cuts. The residual columns are the maximum relative residuals of the linear and cubic models, respectively. The last column is the ratio of the linear-fit SSE using $\eta_{\rm lat}=\ln(m_{\rm lat,1}/m_{\rm lat,2})$ to that using $\eta=\ln r$.

\begin{table}[!htbp]
\caption{Six-ratio fits at $gL_{\rm phys}=8$.  The coefficients, residuals, and SSE ratio are dimensionless; $x$ is a dimensionless lattice-spacing parameter.}
\label{tab:df-odd-fits}
\centering
\scriptsize
\begin{tabular}{@{}c l r r r r r@{}}
\toprule
$x$ & cut & $c_1$ & $c_3$ & $100c_3/c_1$ & max. residual (\%), lin./cubic & $\mathrm{SSE}_{\eta_{\rm lat}}/\mathrm{SSE}_{\eta}$ \\
\midrule
4 & boundary & $-1.275960$ & $0.009970$ & $-0.781$ & $0.252/0.0111$ & 16.618 \\
4 & intra & $-2.894098$ & $0.082275$ & $-2.843$ & $0.981/0.0229$ & 3.384 \\
6.25 & boundary & $-1.121443$ & $0.008691$ & $-0.775$ & $0.253/0.0085$ & 9.447 \\
6.25 & intra & $-2.520336$ & $0.070368$ & $-2.792$ & $0.966/0.0242$ & 2.469 \\
9 & boundary & $-1.019017$ & $0.008436$ & $-0.828$ & $0.272/0.0072$ & 5.961 \\
9 & intra & $-2.267118$ & $0.063666$ & $-2.808$ & $0.971/0.0246$ & 2.022 \\
12.25 & boundary & $-0.945389$ & $0.008422$ & $-0.891$ & $0.294/0.0066$ & 4.241 \\
12.25 & intra & $-2.083003$ & $0.059182$ & $-2.841$ & $0.983/0.0245$ & 1.768 \\
\bottomrule
\end{tabular}
\end{table}

Because $D_F^{\mathrm{conf}}$ is at most $1.292\%$ of $|D_F|$, these fits primarily parameterize $\ln(p_-/p_+)=-2\operatorname{artanh}\mathcal P_F^{(-1)}$. The odd expansion accurately describes the sampled interval but does not establish a constant-exponent power law: the small cubic term gives a resolved scale dependence to the effective exponent. Its coefficients are finite-spacing, finite-length, and cut-dependent descriptors. The table does not establish that fifth and higher odd terms vanish, and the single available six-ratio length $gL_{\rm phys}=8$ does not determine the length dependence of $c_1$ or $c_3$.

\subsection{Normalized-block correction}

For completeness, the relative contribution $|D_F^{\mathrm{conf}}/D_F|$ has median, mean, and maximum values $0.6527\%$, $0.6939\%$, and $1.2916\%$ at the 32 unit-cell boundaries. After reflection-related intra-cell cuts are combined, the corresponding values are $0.0619\%$, $0.0585\%$, and $0.1428\%$. The smaller term is independently resolved over the available bond-dimension increases, but it does not set the main probability response or characterize the complete normalized block spectra.

\FloatBarrier

\subsection{Exact symmetry maps and conserved-sector validation}

Write $H_{Q,F}(r)$ for the Hamiltonian in total sector $(Q,F)$. Flavor swap gives
\begin{equation}
\mathcal XH_{Q,F}(r)\mathcal X^{-1}=H_{Q,-F}\!\left(\frac{1}{r}\right),\qquad
\operatorname{Spec}H_{Q,F}(r)=\operatorname{Spec}H_{Q,-F}\!\left(\frac{1}{r}\right).
\end{equation}
For even chains in the neutral $L_{-1}=0$ internal-link realization, canonical particle--hole reflection $\mathcal Pc_{n,\alpha}\mathcal P^{-1}=(-1)^n c^\dagger_{N_s-1-n,\alpha}$ maps $Q_n\to-Q_{N_s-1-n}$, reflects the hopping and electric sums, preserves the staggered mass term, and yields
\begin{equation}
\mathcal PH_{0,F}(r)\mathcal P^{-1}=H_{0,-F}(r).
\end{equation}
This concise map is distinct from the one-site-translating staggered charge conjugation discussed in Sec.~\ref{sec:model}; it also provides an exact check of the sector construction. The explicit $C$ asymmetry of the open-boundary realization means that these finite-chain calculations do not test the two-vacuum degeneracy of the $C$-broken infinite-volume theory.